# Polarization of the Vaccination Debate on Facebook

Ana Lucía Schmidt[1], Fabiana Zollo[2], Antonio Scala[3] ,Cornelia Betsch[3], Walter Quattrociocchi[2]
[1]IMT Alti Studi Lucca
[2]Ca' Foscari University of Venice
[3]ISC-CNR, Rome Italy
[4]University of Erfurt

## Abstract

### Background

Vaccine hesitancy has been recognized as a major global health threat. Having access to any type of information in social media has been suggested as a potential powerful influence factor to hesitancy. Recent studies in other fields than vaccination show that access to a wide amount of content through the Internet without intermediaries resolved into major segregation of the users in polarized groups. Users select the information adhering to theirs system of beliefs and tend to ignore dissenting information.

### Objectives

In this paper we assess whether there is polarization in Social Media use in the field of vaccination.

### Methods

We perform a thorough quantitative analysis on Facebook analyzing 2.6M users interacting with 298.018 posts over a time span of seven years and 5 months. We used community detection algorithms to automatically detect the emergent communities from the users' activity and to quantify the cohesiveness over time of the communities.

### Results

Our findings show that content consumption about vaccines is dominated by the echo-chamber effect and that polarization increased over years. Communities emerge from the users' consumption habits, i.e. the majority of users only consumes information in favor or against vaccines, not both.

### Conclusion

The existence of echo-chambers may explain why social-media campaigns providing accurate information may have limited reach, may be effective only in sub-groups and might even foment further polarization of opinions. The introduction of dissenting information into a sub-group is disregarded and can have a backfire effect, further reinforcing the existing opinions within the sub-group.

### Keywords

Social media, vaccine hesitancy, network analysis, computational social science, misinformation

# Introduction

Undeterred by the scientific consensus that vaccines are safe and effective, unsubstantiated claims doubting their safety still occur to this day. Perhaps the most famous case is the multiple times disproved [1,2,3] myth that the MMR vaccine causes autism. However, outbreaks and deaths resulting from objections to vaccines continue to happen [4,5], with the anti-vaccination movement gaining media attention as a result. Mandatory vaccination policies only seem to foment the controversy [6].

Since 2013, the World Economic Forum lists massive digital misinformation among the main threats to our societies [7]. Recent studies outline that misinformation spreading is a consequence of the shift of paradigm in the consumption of content induced by the advent of social media. Social media platforms like Facebook or Twitter have created a direct path for users to produce and consume content, reshaping the way people get informed [8-13].

Like for other misinformation campaigns, Facebook provides an ideal medium for the diffusion of anti-vaccination ideas. Users can access a wide amount of information and narratives and selection criteria are biased toward personal viewpoints [14,15,16]. Online users select information adhering to their system of beliefs and tend to ignore dissenting information and to join polarized groups that cooperated to reinforce and frame a shared narrative [17,18,19]. The interaction with content dissenting the shared narrative is mainly ignored or might even foment segregation of users, heated debating and thus bursting polarization of opinions [20]. Such a scenario is not limited just to conspiracy theories, but it is related to all issues that are perceived as critical by the users such as geopolitics and health [21]. This effect allows for the emergence of polarized groups [12], i.e. clusters of users with opposing views that rarely interact with one another.

In this paper we use quantitative analysis to understand the evolution of the debate about vaccines on Facebook, taking into account two opposing views: anti-vaccines and pro-vaccines. Considering the liking and commenting behavior of 2.6M users we study the evolution of the two communities over time, taking into account the number of users, the number of pages, and the cohesiveness of the communities. The analyses confirm the existence of two polarized communities. Additionally, we find evidence that selective exposure plays an pivotal role in the way users consume content online. The two communities display different rates at which the variety of news sources consumed diminishes for increasing levels of user activity, with the anti-vaccines community being more engaged.

# Data Description

## Ethics Statement

The data collection process was carried out using the Facebook Graph API [22], which is publicly available. The pages from which we downloaded data are public Facebook entities and can be accessed by anyone. Users' content contributing to such pages is also public unless users' privacy settings specify otherwise, and in that case their activity is not available to us.

## Data Collection

The dataset was generated through requests to Facebook for pages containing the keywords *vaccine*, *vaccines* or *vaccination* in their name or description. We then filtered the raw Facebook data in order to include only the ones relevant for the study. Inclusion criteria were language (English), a minimum level of activity (at least 10 posts), date of the posts (between 1st January 2010 to 31st May 2017), and relation of the page to vaccination. This last step was essential, as having one of the keywords in the description does not necessarily mean the page's topic is about vaccines. Some examples of those false positive search results are the pages *The Vaccines* (an UK music band), and *Arthur D'vaccine* (a comedian).

From the resulting set of Facebook pages we downloaded all posts as well as all likes and comments made on those posts. Considering the content of the posts made on the pages, all the Facebook pages were also manually classified by two raters into two groups: 145 *pro-vaccines* with 1,388,677 users and 98 *anti-vaccines* with 1,277,170 users. The Cohen's kappa inter-agreement between both raters is 0.966, showing nearly perfect agreement.

A list of Facebook pages with their respective community label and a breakdown of the dataset in numbers of posts, likes, likers, comments, commenters and users can be found in the Appendix.

# Preliminaries and Definitions

The likes (or comments) given by users to the posts of different Facebook pages form a *bipartite network*. The *bipartite network* is formed by a set of users and a set of pages where links only exist between a user and a page if the user liked (or commented) anything on that page.

We can represent the bipartite network with a matrix where each column is a user and each row is a page, and the content of each cell equals 1 if the user liked at least one post of that page. If we multiply the matrix by its transpose, we get the *projection of the bipartite network*. This new matrix will have a row and column for each page, and the content of each cell will represent the number of common users between the 2 pages that define that cell, that is, the number of users who liked any post on both pages. The same method can also be applied considering the matrix formed by the users' comments.

For illustration, Figure 1 visualizes a simplified example of a bipartite network with 5 users and 4 pages and the corresponding projection.

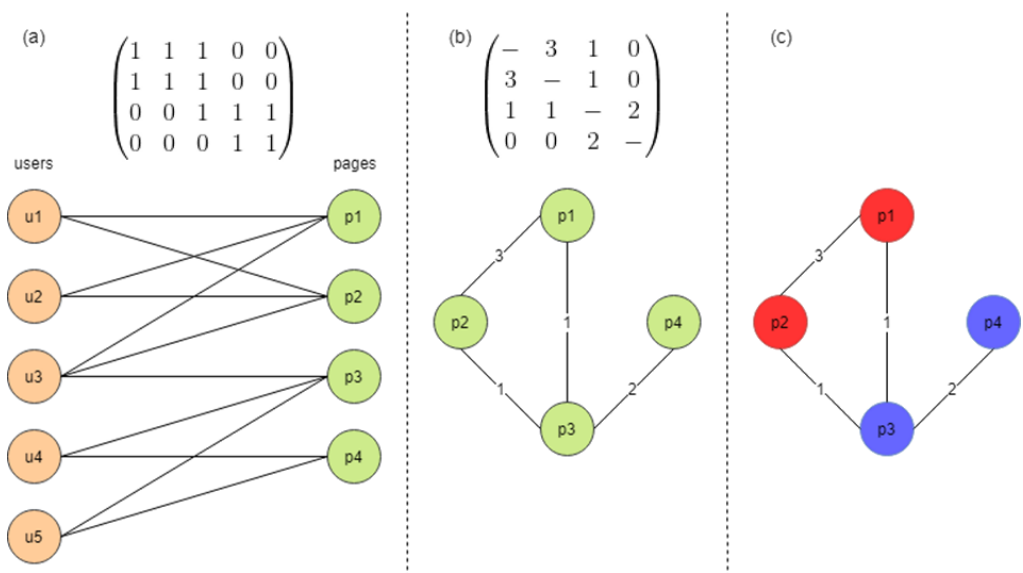

*Figure 1 – (a) Bipartite network with 5 users and 4 pages, the links between them indicate that a user liked a page. (b) The projection of the bipartite network, the weights on the links between the pages show the number of users they have in common. (c) The community structure as detected with the algorithm FastGreedy. Nodes sharing the same color belong to the same community.*

Once we have the network of pages linked by their common users (Figure 1b), we can apply different community detection algorithms to detect group of pages that are strongly connected (Figure 1c). To do this we apply five well known community detection algorithms: FastGreedy[1][23], WalkTrap[2][24], MultiLevel[3][25] and LabelPropagation[4][26]. Different algorithms are used as they allow for unsupervised clustering, i.e., no human intervention, and they each have different approaches to detecting of communities in the networks. To compare the communities detected with the various algorithms we use standard methods that compute the similarity between different community partitions by considering how the algorithms assign the nodes to each community [27]. Due to its speed and its lack of parameters in need of tuning, the FastGreedy algorithm will be the main reference to compare against the partitions resulting from the application of other community detection algorithms.

## Results and Discussion

### Validation of the Community Partition

In order to validate the manual partition of the pages into two communities we generated the projections of the bipartite networks considering the user likes and the user comments. We then applied the community detection algorithms to extract the communities of pages according to the users' behavior and compared those to the expert-based partitioning.

Table 2 shows the comparison between a random partition of the pages, the manual partition, and the FastGreedy partition against those resulting from the different algorithms. We can see that the manual classification matches well against the unsupervised approaches and that the FastGreedy results have a high agreement with the other algorithms. This indicates that the users' behavior generates well defined communities of pages and that these communities are similar to the anti-vaccines and pro-vaccines as manually tagged.

*Table 1 – Validation of the community partition.*

| Graph | Communities | FastGreedy | WalkTrap | MultiLevel | LabelProp. |
|---|---|---|---|---|---|
| **Likes** | **Random** | 0.496 | 0.497 | 0.495 | 0.497 |
| | **Manual** | 0.774 | 0.721 | 0.738 | 0.714 |
| | **FastGreedy** | 1 | 0.935 | 0.950 | 0.901 |
| **Comments** | **Random** | 0.497 | 0.499 | 0.495 | 0.496 |
| | **Manual** | 0.590 | 0.610 | 0.567 | 0.570 |
| | **FastGreedy** | 1 | 0.909 | 0.876 | 0.824 |

Note: We compared a random partition of the pages into communities, the manual classification, and the FastGreedy classification against the community partitions detected with the different community detection algorithms. The values of the comparison range from 0 to 1, where 1 is an exact match and 0 is no match.

---

[1] It optimizes the modularity score in a greedy manner to calculate the communities. The modularity is a benefit function that measures the quality of a particular division of a network into communities. A high modularity score corresponds to a dense connectivity between nodes inside a cluster and sparse connections between clusters. This algorithm takes an agglomerative bottom-up approach: initially each vertex belongs to a separate community and, at each iteration, the communities are merged in a way that yields the largest increase in the current value of modularity.

[2] It exploits the fact that a random walker tends to become trapped in the denser parts of a graph, i.e., in communities.

[3] It uses a multi-level optimization procedure for the modularity score. It takes a bottom-up approach where each vertex initially belongs to a separate community and in each step, unlike FastGreedy, vertices are reassigned to a new community.

[4] It uses a simple approach where each vertex is assigned a unique label, which is updated according to majority voting in the neighboring vertices. Dense node groups quickly reach a consensus on a common label.

Thus, the pages cluster together according to the users' activity. In a next step, we analyzed the polarization of the users.

## Polarization

Assuming that a user u has performed x likes on community C1 and y likes on community C2, we calculate the users' polarization as ρ(u) = (x − y)/(x + y). Thus, a user u for whom ρ(u) = −1 is polarized towards C2, whereas a user whose ρ(u) = 1 is polarized towards C1. We then measure the polarization of all users considering the communities they commented and liked content on. We examine two partitions: the manual classification of pages, pro-vaccine and anti-vaccine, and the two biggest communities as detected with FastGreedy, C1 and C2.

Figure 2 shows the Probability Density Function (PDF) of ρ(u) for all users who have given at least 10 likes in their lifetime. The PDF for the polarization of all users is sharply bi-modal, that is, the majority of the users are either at -1 or at 1. This indicates a strong polarization among the communities, that is, the majority of the users are active either in the pro-vaccines or anti-vaccines community, not both.

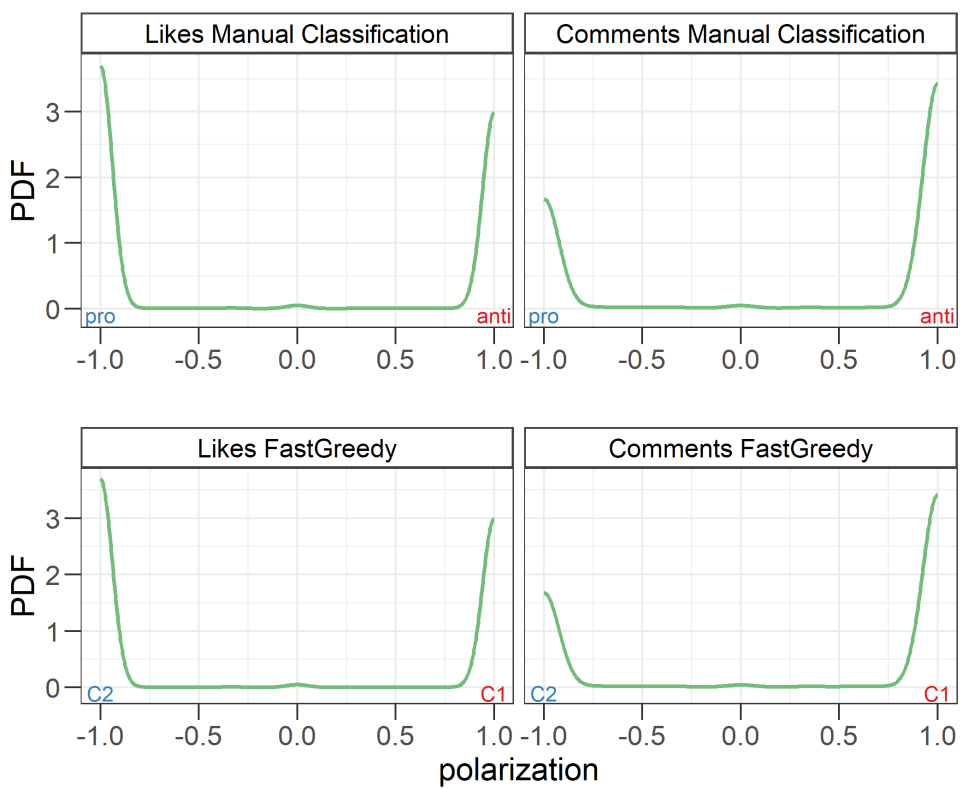

*Figure 2 - Probability Density Function (PDF) of the users' liking (left) and commenting (right) behavior in the manual communities (top) and the 2 largest communities detected with FastGreedy (bottom). The distribution of the users is bimodal for all cases, which indicates a strong polarization among the communities, that is, the majority of the users are active in only one community.*

## Selective Exposure

Facebook users differ in the time they spend with the pages and in how frequently they interact with the pages. The *lifetime* of a user is defined as the period of time where the user started and stopped interacting with the included set of pages. It can be approximated by the time difference between a user's latest and earliest liked post. The total number of likes per user is a good proxy for the user's *activity*, i.e., their level of engagement with the Facebook pages. These two measures provide important insights on how users consume information in each echo chamber as demonstrated in the following analyses.

Figure 3 shows the number of unique pages users from the anti-vaccines (red) and pro-vaccines communities (blue) interact with, considering increasing levels of lifetime and activity for different time windows (yearly left, monthly middle and weekly right panel). For a comparative analysis, we standardized lifetime and activity to range between 0 and 1, both over the entire user set of each community, and the number of pages.

Note that for both communities, users usually interact with a small number of Facebook pages. Longer lifetime and higher levels of activity correspond with less number of pages being consumed. This suggests that more time on Facebook corresponds to a smaller variety of sources being consumed. This is consistent with [12] showing that content consumption on Facebook is dominated by selective exposure and, over time, users personalize their information sources accordingly with their tastes which results in a smaller number of sources.

Pro-vaccine users interact with M = 1.42 pages (SD = 0.79), anti-vaccine users with 2.45 (SD = 2.13). This difference is displayed in Figure 3: users in the anti-vaccines community (red line) consume information from a more diverse set of pages than those in the pro-vaccines community, regardless of the time window considered. Grey shades are 95% CI of the local regression of the data, indicating significant differences between the groups at any time. So while there is a natural tendency of users to confine their activity to a limited set of pages [12], the two communities display different rates of selective exposure. The anti-vaccine community shows more commitment to the consumption of their posts.

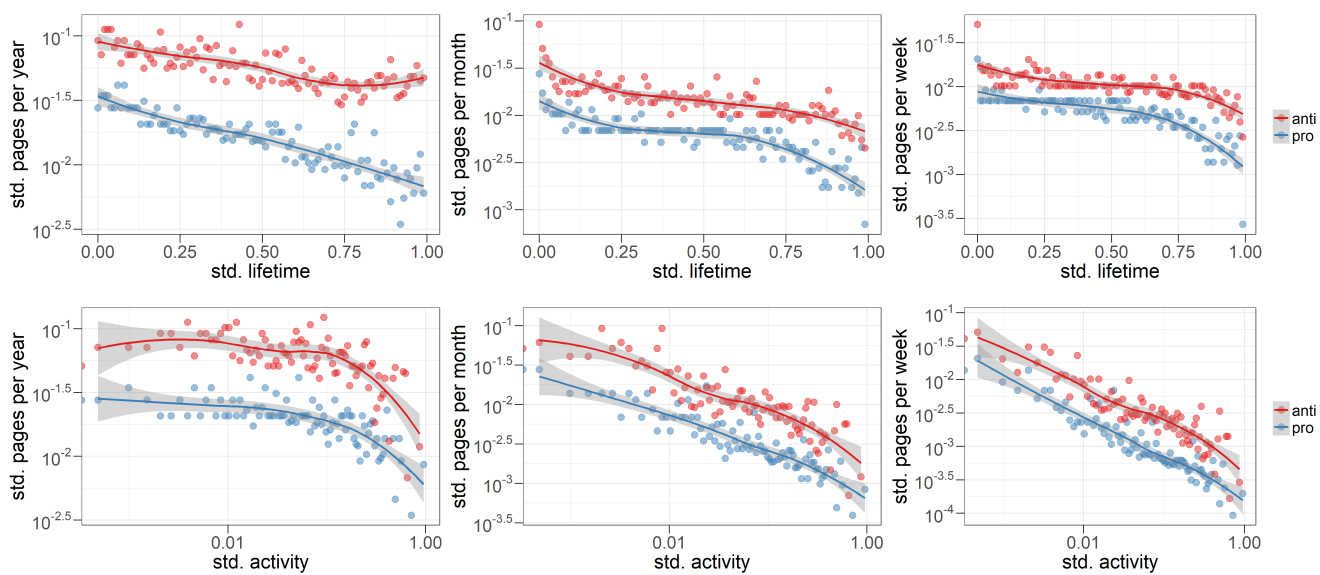

*Figure 3 - Maximum number of unique pages that users with increasing levels of standardized lifetime (top) or standardized activity (bottom) interact with yearly (left), monthly (middle) and weekly (right) for each community. Users' lifetime corresponds to the standardized time difference between their latest and earliest liked post. Users' activity corresponds to the standardized number of likes given in their lifetime. Users display a tendency to like less pages when their lifetime and activity increases. The users who interact with the anti-vaccines community also consume a larger variety of pages than the pro-vaccines users. Grey shades are 95% CI of the fitted curve, indicating significant differences between the groups at any time.*

## Growth of the Communities over Time

We also analyzed the growth of the two communities over time, considering the variety of pages and the number of users that interact with them. Figures 4 shows the evolution of the communities over the years in quarterly increments.

The left panel plots the number of active pages in each community. We define a page as active in a specific quarter if it made a post (bottom panel), received a like (middle) or comment in that period (upper panel).

Overall, the number of active pages in both communities increases at similar rates, with slight variations when we consider the different types of action that marks a page as active. If we use the pages' posting activity or the likes they received to determine whether they were active in a given quarter, we can see that, from 2013, the pro-vaccine community consistently tends to show a higher number of active pages than the anti-vaccine community (interaction effect in a MANOVA with sentiment (pro, anti) and time (until 2012Q4 vs. following) as factors and posts and likes as dependent variables $F(2,55) = 2.708$, $p = 0.076$; $eta^2 = 0.09$; both main effects are highly significant). On the other hand, if we focus on the comments, the anti-vaccines community shows a boost in activity starting in 2015 (interaction effect in an ANOVA with sentiment (pro, anti) and time (until 2014Q4 vs. following) as factors and comments as dependent variable $F(1,56) = 5.053$, $p = 0.029$; $eta^2 = 0.08$; both main effects are significant).

The right panel plots the number of active users in each community. We define users as active if they gave a like (or comment) to any page of that community in the given quarter. The plot shows that while both communities gain users throughout the entire period, the anti-vaccines community has, until the end of 2015 and beginning of 2016, more active users than the pro-vaccines community. After that, this relation reverses (interaction effect in a MANOVA with sentiment (pro, anti) and time (until 2015Q4 vs. following) as factors and comments and likes as dependent variables $F(2,55) = 12.218$, $p < 0.001$; $eta^2 = 0.31$; both main effects are highly significant).

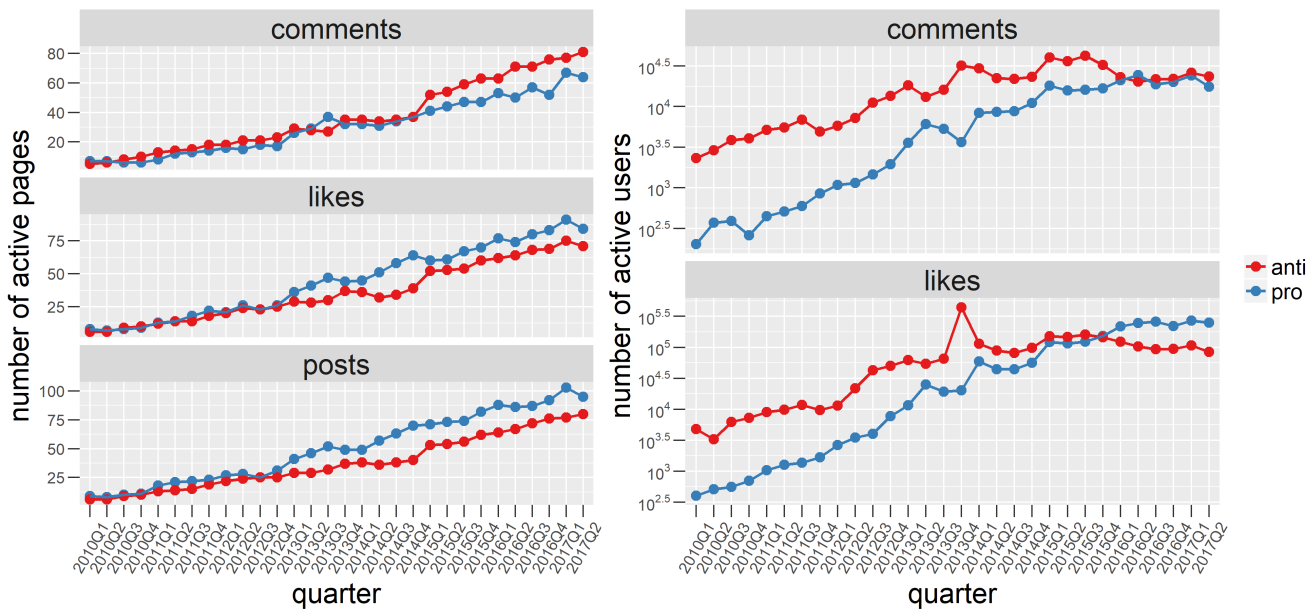

*Figure 4 – Number of active pages (left) and users (right) in each community. We define a page as active in a specific quarter if it made a post (bottom panel), received a like (middle panel) or comment (upper panel) in that period. We define a user as active in a community on a given quarter, if they gave a like (bottom panel) or comment (top panel) to any page of that community in that time.*

Another important factor to consider is the cohesiveness of the pro-vaccines and anti-vaccines communities. In order to analyze whether the growth of the communities depends on the emergence of isolated pages or whether it grows steadily, we split the projections of the bipartite likes and comments graph by the community of the pages. This results in 4 sub-graphs, each containing the pages of one community, pro-vaccines or anti-vaccines, and the common users that linked them considering the likes or the comments. We can then calculate the fragmentation of each community by applying the community detection algorithms and obtaining their partition.

Figure 5 shows the number of pages of the biggest sub-community of the anti-vaccines (left) or pro-vaccines communities (right) in a given quarter, that is, the largest connected component found with the different community detection algorithms. The black line represents the total number of pages in the sub-graphs in that quarter. It marks the maximum possible size for the largest connected component to take in that moment in time. The closer the size of the largest connected component is to the total number of pages in the system, the more tightly linked that community is in that moment in time.

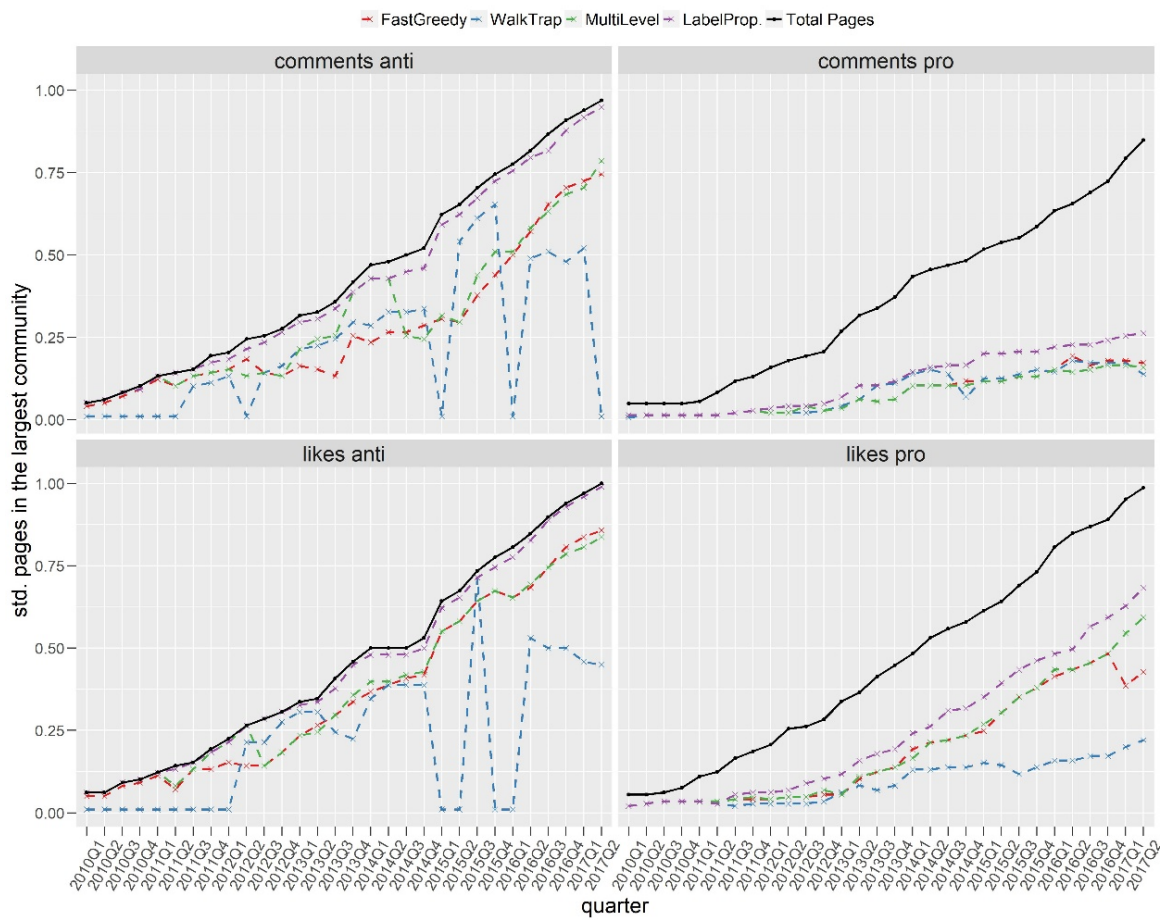

*Figure 5 - Size of the largest connected component within the set of pages tagged as anti-vaccines and pro-vaccines over time, considering various community detection algorithms. The black line represents the total number of pages over time in the anti-vaccines and pro-vaccines communities, that is, the maximum possible size for the largest connected component in that moment in time. The graph shows that the anti-vaccines community grows cohesively, with the new pages joining the already existing group of pages, while the pro-vaccines community grows in a more fragmented, independent way.*

The plots show that in the anti-vaccines community the number of pages in the largest component remains close to the total number of pages in the system. In the case of the pro-vaccines sub-graphs, however, the size of the largest community does not increase closely with the number of pages in the system. This signifies that the anti-vaccines community grows in a more cohesive manner, with pages tightly linked by their users' activity, while the pro-vaccines community is more fragmented.

# Discussion

By means of quantitative analysis of Facebook likes and comments we validated the existence of two opposing narratives regarding the vaccination debate on Facebook. We show that the communities emerge from the users' consumption habits and that users are highly polarized, that is, the majority of users only consumes and produces information in favor or against vaccines, not both.

We also showed that both narratives are subjected to selective exposure, and that the more active a user is on Facebook the smaller is the variety of sources they tend to consume. We note, however, that the users from the anti-vaccination community consume more sources compared to the pro-vaccine users. This is consistent with the results of previous studies [14] that show that people in conspiracy-like groups show higher engagement with the community.

We also analyzed the communities' evolution over time. While the pro-vaccine pages are generally more active, the anti-vaccine pages concentrate the majority of the debate, receiving more comments from users. We show that the anti-vaccine community had a more active user base until the end of 2015, where the activity seems to stall. This matches with the outbreak of measles at Disneyland [4], which put the anti-vaccination movement in the spotlight and gained the attention of mainstream media [28-34]. Further studies are needed to determine the reason for this stagnation.

Finally, we show that while both narratives have gained attention on Facebook over time, anti-vaccine pages display a more cohesive growth (i.e. pages are liked by the same people), while the pro-vaccine pages seem to grow in a highly fragmented fashion (i.e. pages are liked by different people).

## Limitations

The data collection process was done the 5th of June 2017 and represents a snapshot of the pages, posts, comments and likes available at the time. Pages, posts, likes and comments that were made in the downloaded period (1st January 2010 to 31st May 2017) and were removed before the download date are not present in the dataset. The data only includes the likes and comments by users whose privacy settings allowed public access to their activity on public pages on the download date.

# Conclusions

Facebook allows echo-chambers to emerge, in which pro- and anti-vaccination attitudes polarize the users. Social media campaigns that advocate for vaccination and provide accurate information should expect to only reach pro-vaccination users as there is nearly no interaction between the groups. Overall, social media seem to be a powerful promoter of different sentiments about vaccination and therefore it is likely that it contributes to vaccine hesitancy.

# Appendix

*Table 2 - Dataset Description.*

|  | Anti-vaccines | Pro-vaccines |
|---|---|---|
| **Pages** | 98 | 145 |
| **Posts** | 189,759 | 108,259 |
| **Likes** | 12,696,440 | 11,459,295 |
| **Likers** | 1,145,650 | 1,325,511 |
| **Comments** | 1,351,839 | 749,209 |
| **Commenters** | 271,598 | 146,196 |

| Users | 1,277,170 | 1,388,677 |

Note: The posts, likes and comments are considered pro or anti vaccines if they were made on a page classified as such. Likers is the number of unique users who have given at least one like to the community. Commenters is the unique number of users who have given at least one comment to the community. Users is the number people who have given at least a like or a comment to the community.